\documentclass[twocolumn,floatfix,prb,aps,showpacs]{revtex4-1}
\usepackage{graphicx,amsmath,amssymb}

\begin{document}

\newcommand{\CORR}[1]{\textbf{#1}}

\title{Change in the character of quasiparticles without gap collapse in a model of fractional quantum Hall effect}
\author{Csaba T\H oke$^{1,2}$ and Jainendra K.~Jain$^{3}$}
\affiliation{$^{1}$Physics Department, Lancaster University, LA1 4YB, Lancaster, United Kingdom}
\affiliation{$^{2}$Institute of Physics, University of P\'ecs, 7624 P\'ecs, Hungary}
\affiliation{$^{3}$Department of Physics, 104 Davey Laboratory, Pennsylvania State University, University Park, Pennsylvania 16802, USA}
\date{\today}

\begin{abstract}
It is commonly assumed in the studies of the fractional quantum Hall effect that
the physics of a fractional quantum Hall state, in particular the character of its excitations, is invariant under a continuous deformation of the Hamiltonian during which the gap does not close.  We show in this article that, 
at least for finite systems, as the interaction is changed from a model three body interaction to Coulomb, the ground state at filling factor $\nu=2/5$ evolves continuously from the so-called Gaffnian wave function to the composite fermion wave function, but the quasiholes alter their character in a nonperturbative manner. This is attributed to the fact that the Coulomb interaction opens a gap in the Gaffnian quasihole sector, pushing many of the states to very high energies. Interestingly, the states below the gap are found to have a one-to-one correspondence with the composite fermion theory, suggesting that the Gaffnian model contains composite fermions, and that the Gaffnian quasiholes are unstable to the formation of composite fermions when a two-body interaction term is switched on. General implications of this study are discussed.
\end{abstract}

\pacs{73.43.Cd, 71.10.Pm}

\maketitle

\section{Introduction}

Topological properties, by definition, are invariant under a continuous deformation of parameters until a phase boundary is encountered.  It is believed that two fractional quantum Hall states that are adiabatically connected, that is, are related by a deformation of the Hamiltonian during which the gap does not close, have identical topological properties.  To the extent the character of the excitations is determined by the topology of the state, this implies that the excitations also evolves adiabatically.  However, there is no reason why the excitations cannot change their character in a fundamental manner through level crossings and gap openings in the excitation spectrum even as the ground state evolves adiabatically. Such examples indeed exist in other contexts: for Bardeen-Cooper-Schrieffer to Bose-Einstein condensate crossover of a superconductor, the low-energy excitations are fermionic quasiparticles in the Bardeen-Cooper-Schrieffer limit but bosonic collective modes in the Bose-Einstein condensate limit. Some examples of such phase transitions in the fractional quantum Hall effect\cite{Tsui} (FQHE) are given at the end of this article.  This paper concerns a model in which the gap does not close during a variation in the interaction but the nature of the quasiholes changes in a qualitative manner.

The FQHE in the lowest Landau level is explained by the composite fermion (CF) theory.\cite{Jain,Dev}  Although not widely appreciated, the topological character of the FQHE is  encoded in the very formation of composite fermions, which are topological particles by virtue of having quantized vortices as one of their constituents.  The topology of composite fermions reveals itself most directly through an effective magnetic field $B^*=B-2p\rho\phi_0$ for composite fermions, as well as their $\Lambda$ levels ($\Lambda$ levels are the kinetic energy levels of composite fermions, analogous to the Landau levels of electrons in the effective magnetic field),
which are directly responsible for most of the FQHE phenomenology (here, $\phi_0\equiv hc/e$ is called the flux quantum, $\rho$ is the particle density, and $2p$ is the number of vortices bound to electrons). The prominent sequences of fractions at $\nu=n/(2pn\pm 1)$ are explained as the integral quantum Hall effect (IQHE) of composite fermions, and the $\nu=1/2$ state as the Fermi sea of composite fermions\cite{HLR} in $B^*=0$. (The electron filling factor is defined as $\nu=\rho hc/eB$.) The effective magnetic field $B^*$ has also been measured in several geometric experiments.\cite{FSexp}  Theoretically, the effective magnetic field and $\Lambda$ levels have been confirmed by verifying that (i) the low-lying energy levels of the interacting electron system at an {\em arbitrary} $B$ have a one-to-one correspondence with the low-lying energy levels of noninteracting fermions at $B^*$, and (ii)  the wave functions of interacting electrons at $B$ are closely related to noninteracting fermions at $B^*$ (through composite fermionization).  
Of relevance to this article are states away 
from the special fillings $n/(2n\pm 1)$, which are described in terms of states in which the topmost $\Lambda$ level contains only a few composite fermions [when $\nu>n/(2n\pm 1)$] or a few holes [when $\nu<n/(2n\pm 1)$]; these are sometimes called CF-quasiparticles or CF-quasiholes.  The energy level counting for the states containing many CF-quasiparticles or CF-quasiholes can be obtained very simply by modeling them as noninteracting fermions at an effective magnetic field; the residual interaction, however, will convert the exact degeneracy into a quasi-degeneracy.

The topological character of the FQHE states is
believed also to manifest itself through adiabatic braiding
properties of far separated quasiparticles or quasiholes,
which are interpreted in terms of fractional braiding statistics.\cite{fracstats}
This concept was first introduced by Halperin \cite{Halperin84},
and can also be derived within the CF theory \cite{cffracstats,Jainbook}.

Other wave functions have been constructed with different topological content. Many of these wave functions\cite{theory5p2,parafermion,multicomp,Gaffnian} represent unique, maximum density zero-energy ground states of model Hamiltonians that do not have a two-body interaction but impose a penalty when three or more particles occupy a finite number of relative angular momentum states. Special cases have been named Pfaffian,\cite{theory5p2} parafermion,\cite{parafermion} Haffnian,\cite{multicomp} or Gaffnian,\cite{Gaffnian} but all of these wave functions can be expressed conveniently as fully antisymmetrized correlated multicomponent wave functions, and also have been interpretated as certain correlation functions of appropriately chosen conformal field theories. The so-called ``Jack" wave functions, are obtained from a root state with the help of certain squeezing rules.\cite{Jack}  Some of these states have been argued to support excitations that satisfy nonabelian braiding statistics.\cite{nonabelian}

One may ask, what is the motivation for constructing new wave functions? (They may turn out to be useful in other contexts, but we confine our attention to the FQHE here.) For the lowest Landau level physics, one may hope to discover a principle for the FQHE that is more fundamental and / or more accurate than that of composite fermions. Given the successes of the CF theory for the lowest Landau level phenomenology, and the fact that composite fermions have been directly observed, it would seem sensible that any new principle must recover, at the least, composite fermions and their physics, such as effective magnetic field, $\Lambda$ levels, unification of the FQHE and the IQHE, etc.

In our opinion, the primary motivation for seeking new FQHE wave functions comes from FQHE in {\em higher} Landau levels \cite{higherexp} which is not as well-described by the CF theory as the lowest Landau level FQHE.  A well-studied case is the FQHE state at 5/2, for which the most widely employed model considers a Pfaffian wave function proposed by Moore and Read.\cite{theory5p2}  While this is a paired state of composite fermions, thus also a part of the CF paradigm, it is also the exact ground state of a model three body interaction which entails an energy cost when three particles occupy the lowest allowed angular momentum state, but no energy cost when two particles approach one another.  (This will be referred to as the ``Pfaffian model Hamiltonian" below.) The Pfaffian wave function has been shown to have a reasonably good overlap with the exact Coulomb ground state,\cite{pfaffoverlap} which has motivated other wave functions that are exact ground states of generalized multi-particle interactions.

\begin{table}[htb]
\begin{center}
\begin{tabular}{c|c|c|c|c|c}
\hline\hline
$N$ & $|\langle\Psi^{\text{CF}}|\Psi^{\text{C}}\rangle|^2$ & $|\langle\Psi^{\text{G}}|\Psi^{\text{C}}\rangle|^2$ &
$|\langle\Psi^{\text{G}}|\Psi^{\text{CF}}\rangle|^2$ & $D_{L_z=0}$ & $D_{L=0}$ \\
\hline
6  & 0.9993 & 0.976 & 0.980 & 58 & 3\\
8  & 0.9986 & 0.955 & 0.962 & 910 & 8 \\
10 & 0.9956  & 0.943 & 0.954 & 16660 & 52\\
\hline\hline
\end{tabular}
\end{center}
\caption{\label{overlap}
Overlaps between the exact Coulomb ground state $\Psi^{\text{C}}$, the Gaffnian state $\Psi^{\text{G}}$, and the composite fermion state $\Psi^{\text{CF}}$ at $\nu=2/5$
in the spherical geometry.
Also shown are the dimensionalities of the Hilbert space in the $L_z=0$ sector ($D_{L_z=0}$) and the $L=0$ sector ($D_{L=0}$).
Some of the overlaps were given previously ($|\langle\Psi^{\text{CF}}|\Psi^{\text{C}}\rangle|^2$ in Ref.~\onlinecite{Dev}; $|\langle\Psi^{\text{G}}|\Psi^{\text{CF}}\rangle|^2$ in Ref.~\onlinecite{multicomp} for up to 14 particles; $|\langle\Psi^{\text{G}}|\Psi^{\text{C}}\rangle|^2$ for $N=10$ in Ref.~\onlinecite{Gaffnian}), but are included here for completeness.
}
\end{table}

The present study is motivated by the so-called ``Gaffnian" wave function\cite{Gaffnian,multicomp} for the 2/5 state, which is the exact ground state of the ``Gaffnian model Hamiltonian" which contains no two particle interactions, but the three particle interactions act in the lowest two relevant angular momenta (as opposed to the lowest relevant angular momentum for the Pfaffian model Hamiltonian). What makes this state particularly interesting is that, as shown in Table \ref{overlap}, it has a reasonably high overlap with the exact Coulomb ground state at 2/5 (in the {\em lowest} Landau level) as well as with the CF wave function.  Analogous wave function for the 3/5 state also has a high overlap with the CF wave function. \cite{multicomp}

This raises the interesting general question: what are the criteria for determining if a given approach is valid for a certain fraction? The validity of the CF theory for the 2/5 state is not in doubt, because it is not just a theory of 2/5 but has numerous other consequences which have been tested and confirmed in excruciating detail, both theoretically and experimentally.  However, one can ask how one may ascertain the validity of a model, such as the Gaffnian model for the 2/5 FQHE, without appealing to the broader phenomenology, because one may encounter situations where a satisfactory understanding of the broader phenomenology has not yet been achieved (as is the case for the second Landau level FQHE). The Gaffnian is less accurate than the CF wave function, but it is not so far from the Coulomb solution as to clearly rule itself out. Were it not for the CF wave function, the Gaffnian would look quite good.

One might be tempted to conclude that the Gaffnian model is also valid and describes the same physics as the CF wave function. The physics of the two models, however, turns out to be qualitatively (topologically) distinct, as clarified by a consideration of their quasiholes and quasiparticles.  In the CF theory, the FQHE state at $\nu=2/5$ maps into the $\nu^*=2$ IQHE of composite fermions, and the ground state wave function with two filled $\Lambda$ levels.
The low-lying states $q$ flux quanta away are described by either $2q$ CF-quasiparticles in the third $\Lambda$ level or as many CF-quasiholes in the second $\Lambda$ level; this makes a definite prediction for the quantum numbers of the low-energy states and their wave functions.  The Gaffnian model  also makes definite prediction about the quantum numbers of the low-lying states and their wave functions on the quasihole side; these are given by the states that have zero-energy for the Gaffnian model Hamiltonian. The energy level counting in the presence of several quasiparticles or quasiholes is in general different in the two approaches, indicating fundamentally distinct topological structures. We are thus faced with a situation where two models produce good wave functions for the 2/5 ground state, but no more than one of them can be correct for the Coulomb interaction.

We present in this paper a careful study of states containing many ``quasiholes" obtained when additional magnetic flux is introduced in the context of the 2/5 state, by testing the predictions of the Gaffnian and the CF models against exact results for finite systems. Our results show that a consideration of quasiholes allows one to distinguish between the two models.

One criticism that can be leveled against our study is that the numerical systems accessible to us are too small to capture the braiding properties, and therefore inconclusive.  While the numerical systems may well be inconclusive, it should be noted that there are two different issues at stake here.  A proper evaluation of the braiding properties would indeed require very large systems, especially in cases where one needs to consider four well-separated quasiparticles or quasiholes.\cite{ssc,baraban}  However, there is no fundamental reason why the Gaffnian model should not give the correct level counting (i.e., quantum numbers of states in the low-energy band) and accurate wave functions even for small systems containing several quasiparticles or quasiholes.  An example in case is the CF theory for the quasiparticles and quasiholes of the FQHE states at the $n/(2n+ 1)$. While an evaluation of the braiding statistics of the CF-quasiparticles or CF-quasiholes indeed requires large systems,\cite{cffracstats} the composite fermion theory gives an exceedingly accurate account of the states containing many quasiparticles and quasiholes even for very small systems, both through level counting and their microscopic wave functions.

We note that a recent paper by Regnault, Bernevig, and Haldane\cite{Regnault} also investigates the issue of how one can discriminate between the CF and Gaffnian wave functions at 2/5.  They compute the topological entanglement to find that the CF theory better agrees with the Coulomb results.  That work considers the ground state wave functions, whereas we consider below the behavior as the system moves away from the ground state; however, the work in Ref.~\onlinecite{Regnault} is related to our work in spirit, because the topological entanglement of a state is a probe into the character of its edge excitations.

The rest of the article is arranged as follows.
In Sec.~\ref{models}, we review the Gaffnian and the composite fermion models.
Sec.~\ref{incomp} argues, through exact diagonalization of the Gaffnian model, that incompressibility at 2/5 cannot rule out for in some range of the parameters
of the Gaffnian model interaction. Sec.~\ref{testing} deals with quasiholes, and also shows that the nature of the excitations changes in a fundamental manner as one interpolates between the Coulomb and the Gaffnian interaction. In Sec.~\ref{contain}, we discuss the extent to which the composite fermion quasiholes are contained in the Gaffnian zero-energy band, which, in general, contains more states.
The paper is concluded in Sec.~\ref{discussion} with a discussion of the broader implications of our findings.

\section{Models}
\label{models}

The numerical work reported in this article is performed in the spherical geometry, in which electrons move on the
surface of a sphere and a radial magnetic field is produced by a magnetic monopole of strength $Q$ at the center.\cite{spherical}
Here, $2Q\phi_0$ is the magnetic flux through the surface of the sphere, and
$2Q$ is an integer according to Dirac's quantization condition.\cite{Dirac31}
The single particle states are monopole harmonics $Y_{Qlm}$, where $l=Q+n$ is the angular momentum with $n=0,1,\ldots$ being the Landau level (LL)  index,
$m=-l,-l+1,\ldots,l$ is the $z$-component of angular momentum.
The Coulomb interaction is evaluated with the chord distance.\cite{Fano}

\subsection{Gaffnian model}

The Gaffnian model\cite{Gaffnian} is defined in terms of a generalized projection Hamiltonian, which in the spherical geometry takes the form
\begin{equation}
\label{hami}
\hat H^{\text{G}}=A\sum_{i<j<k}P^{(3)}_{ijk}(3Q-3) + B\sum_{i<j<k}P^{(3)}_{ijk}(3Q-5),
\end{equation}
where $A$ and $B$ are positive constants, and $P^{(3)}_{ijk}(L)$ projects the state of particles $i,j,k$ to the total angular momentum $L$ subspace.
While $\hat H^{\text{G}}$ does not contain any two-body interactions, it penalizes states of three electrons in their smallest two possible angular momentum states.
(The value $L=3Q-4$ of the total angular momentum is excluded by symmetry, and $L>3Q-3$ by the Pauli principle.)

This model has a unique zero-energy ground state at $2Q=5N/2-4$, which has been named ``Gaffnian" state.\cite{Gaffnian} It is given by
\begin{multline}
\Psi^{\text{G}}=\Phi_1 S\left[  \prod_{j<k} [(z_j-z_k)^2 (w_j-w_k)^2]\right.
\times\\
\left.\times\prod_{j,k} (z_j-w_k)\prod_j{1\over (z_j-w_j)} \right]
\end{multline}
where $\Phi_1$ is the wave function of a filled Landau level; electrons have been separated into two clusters $\{z_j\}$ and $\{w_j\}$, $j,k=1,\cdots, N/2$; and $S$ is the symmetrization operator. This is the maximum density zero-energy ground state of the Hamiltonian in Eq.~(\ref{hami}).

When the number of flux quanta is increased by $n$ units, $2n$ quasiholes are generated in the Gaffnian model, reflecting the paired nature of this state.
These states also have zero-energy for the model interaction $\hat H^{\text{G}}$, and a basis for the wave functions can be obtained by inserting inside the symmetrized part the factor
\begin{equation}
\prod_{j=1}^{N/2}\prod_{\alpha=1}^{n}(z_j-Z_\alpha)(w_j-Z_{n+\alpha})
\end{equation}
where $Z_{\alpha}$ are the quasihole positions.  We will refer to these states (denoted $\Psi^{\text{G}}_{2-\text{qh}}$, $\Psi^{\text{G}}_{4-\text{qh}}$, etc) as the ``Gaffnian quasihole sector''.  Several wave functions can be created in this manner, because of the freedom related with which of the $n$ quasihole positions are associated with the $\{z_j\}$'s (with the remaining being associated with $\{ w_j\}$'s), but, as shown in Ref.~\onlinecite{Gaffnian}, these are not all linearly independent.  We will produce an independent basis by numerical diagonalization of the Gaffnian Hamiltonian, which directly generates the Gaffnian quasihole states in the angular momentum basis. The angular momenta of the states in the Gaffnian quasihole sector are enumerated in Tables \ref{counting} and \ref{countingodd} for even and odd numbers of quasiholes. (The latter requires an odd number of electrons.) It has been argued in Ref.~\onlinecite{Gaffnian} that, should the gap remain finite in the thermodynamic limit (see Sec.~\ref{incomp} for further discussion of this issue),
the Gaffnian zero modes likely obey semionic exclusion statistics and nonabelian braiding statistics.

While one expects quasiparticles at flux values $2Q<5N/2-4$, their angular momentum counting is not known analytically due to
the absence of an exact solution for Gaffnian quasiparticles.  It is therefore uncertain how many low-energy states of  $\hat H^{\text{G}}$ should
be considered as elementary quasiparticle excitations.

\begin{table}[htb]
\begin{center}
\begin{tabular}{c|c|c|l|l}
\hline\hline
$N$ & $2Q$ & state & CF band & Gaffnian band \\
\hline
6 & 9  & 4 qp's & $0$     &  Not known \\
6 & 10 & 2 qp's & $1,3$   &  Not known \\
6 & 12 & 2 qh's & $1,3$   & $1,3$        \\
6 & 13 & 4 qh's & $0,2,4$ & $0^2,2^2,3,4^2,6$ \\
\hline
8 & 14 & 4 qp's & $2$     &   Not known \\
8 & 15 & 2 qp's & $0,2,4$   &  Not known \\
8 & 17 & 2 qh's & $0,2,4$   & $0,2,4$        \\
8 & 18 & 4 qh's & $0,2,3,4,6$ & $0^2,2^3,3,4^3,5,6^2,8$ \\
\hline
10 & 19 & 4 qp's & $0,2,4$           & Not known \\
10 & 20 & 2 qp's & $1,3,5$           & Not known \\
10 & 22 & 2 qh's & $1,3,5$           & $1,3,5$        \\
10 & 23 & 4 qh's & $0,2^2,4^2,5,6,8$ & $0^2,2^4,3,4^4,5^2,6^3,7,8^2,10$ \\
\hline\hline
\end{tabular}
\end{center}
\caption{\label{counting}
Angular momenta of the low-energy excitations of the composite fermion model and the Gaffnian model
at flux values ranging from $2Q=5N/2-6$ (four quasiparticles) to $2Q=5N/2-2$ (four quasiholes) for $N$ even.
}
\end{table}

\begin{table}[htb]
\begin{center}
\begin{tabular}{c|c|c|l|l}
\hline\hline
$N$ & $2Q$ & state & CF band & Gaffnian band \\
\hline
7 & 12 & 3 qp's & $1,3$         &  not known \\
7 & 13 & 1 qp   & $\frac{5}{2}$         &  not known \\
7 & 14 & 1 qh   & $2$           & $2$        \\
7 & 15 & 3 qh's & $\frac{3}{2},\frac{5}{2},\frac{9}{2}$ & $\frac{3}{2},\frac{5}{2},\frac{9}{2}$ \\
\hline
9 & 17 & 3 qp's & $\frac{3}{2},\frac{5}{2},\frac{9}{2}$ &   not known \\
9 & 18 & 1 qp   & $3$           &  not known \\
9 & 19 & 1 qh   & $\frac{5}{2}$         & $\frac{5}{2}$        \\
9 & 20 & 3 qh's & $0,2,3,4,6$   & $0,2,3,4,6$ \\
\hline
11 & 22 & 3 qp's & $0,2,3,4,6$   &   not known \\
11 & 23 & 1 qp   & $\frac{7}{2}$         &  not known \\
11 & 24 & 1 qh   & $3$           & $3$        \\
11 & 25 & 3 qh's & $\frac{3}{2},\frac{5}{2},\frac{7}{2},\frac{9}{2},\frac{11}{2},\frac{13}{2}$   & $\frac{3}{2},\frac{5}{2},\frac{7}{2},\frac{9}{2},\frac{11}{2},\frac{13}{2}$ \\
\hline\hline
\end{tabular}
\end{center}
\caption{\label{countingodd}
Angular momenta of the low-energy excitations of the composite fermion model and the Gaffnian model
at flux values ranging from $2Q=5N/2-11/2$ (three quasiparticles) to $2Q=5N/2-5/2$ (three quasiholes) for $N$ odd.
}
\end{table}

\subsection{Composite fermions}

The CF ground state at $\nu=2/5$ is\cite{Jain}
\begin{equation}
\Psi^{\text{CF}} = \mathcal P_{\text{LLL}}\Phi_1^{2}\Phi_2,
\end{equation}
where $\Phi_2$ is the Slater determinant with $N$ fermions completely filling two $\Lambda$ levels.
The Jastrow factor $\Phi_1^{2}=\prod_{i<j}(u_iv_j-u_jv_i)^2$ [with $u=\cos\left(\theta/2\right)e^{-i\phi/2}$ and $v=\sin\left(\theta/2\right)e^{i\phi/2}$]
attaches two vortices to each fermion to convert it into a composite fermion; $\mathcal P_{\text{LLL}}$ projects the wave function on its right into the lowest Landau level.\cite{projection}
The ground state occurs at $2Q=5N/2-4$, which is identical to the flux for the Gaffnian state.
If $2Q$ is increased by $n$ (integer), there is room for $2n$ CF-quasiholes in the second $\Lambda$ level of composite fermions, whereas a reduction of $2Q$ by
$n$ results in $2n$ CF-quasiparticles in the third $\Lambda$ level.
These states are denoted $\Psi^{\text{CF}}_{2-\text{qh}}$, $\Psi^{\text{CF}}_{4-\text{qh}}$, $\Psi^{\text{CF}}_{2-\text{qp}}$, $\Psi^{\text{CF}}_{4-\text{qp}}$, etc.
The angular momenta of these states are given in Table \ref{counting}.
States with an odd number of CF-quasiparticles / CF-quasiholes occur at odd $N$, with quantum numbers predicted by the CF theory enumerated in Table \ref{countingodd}.
Notice that for one, two, and three quasiholes, the Gaffnian interaction predicts the same angular momentum distribution as the CF theory,
but for four or more quasiholes the Gaffnian model produces a significantly greater number of states.

The construction of the basis functions for composite fermions, which are related to the basis functions at the corresponding noninteracting electron system, has been discussed in detail in the past and will not be repeated here. We refer the reader to the literature for the treatment of lowest Landau level projection and CF diagonalization.\cite{projection,theory4p11}

\section{Incompressibility}
\label{incomp}

At least for finite systems, exact diagonalization studies on the Gaffnian model produce an incompressible state with a clear gap.  For these systems, the Gaffnian model produces a qualitatively different structure for the quasiholes than the CF model, and it is valid to ask how the two are related. That is the objective of our calculations below.

It has been suggested in Refs.~\onlinecite{Gaffnian} and \onlinecite{unitary} that the Gaffnian state is not a gapped state in the thermodynamic limit, based on its identification to a critical conformal field theory, the edge theory of which is nonunitary and therefore unphysical.
This argument, however, is not mathematically rigorous at this stage.
From a microscopic perspective, where this state is viewed as the exact maximum density solution of a well-defined Hamiltonian, there is no fundamental reason to doubt the presence of a gap.  As a function of the filling factor, the energy is zero for $\nu\leq 2/5$ but nonzero for $\nu>2/5$ indicating the possibility of a discontinuous change in the chemical potential.

\begin{figure}[!htbp]
\begin{center}
\includegraphics[width=\columnwidth,keepaspectratio]{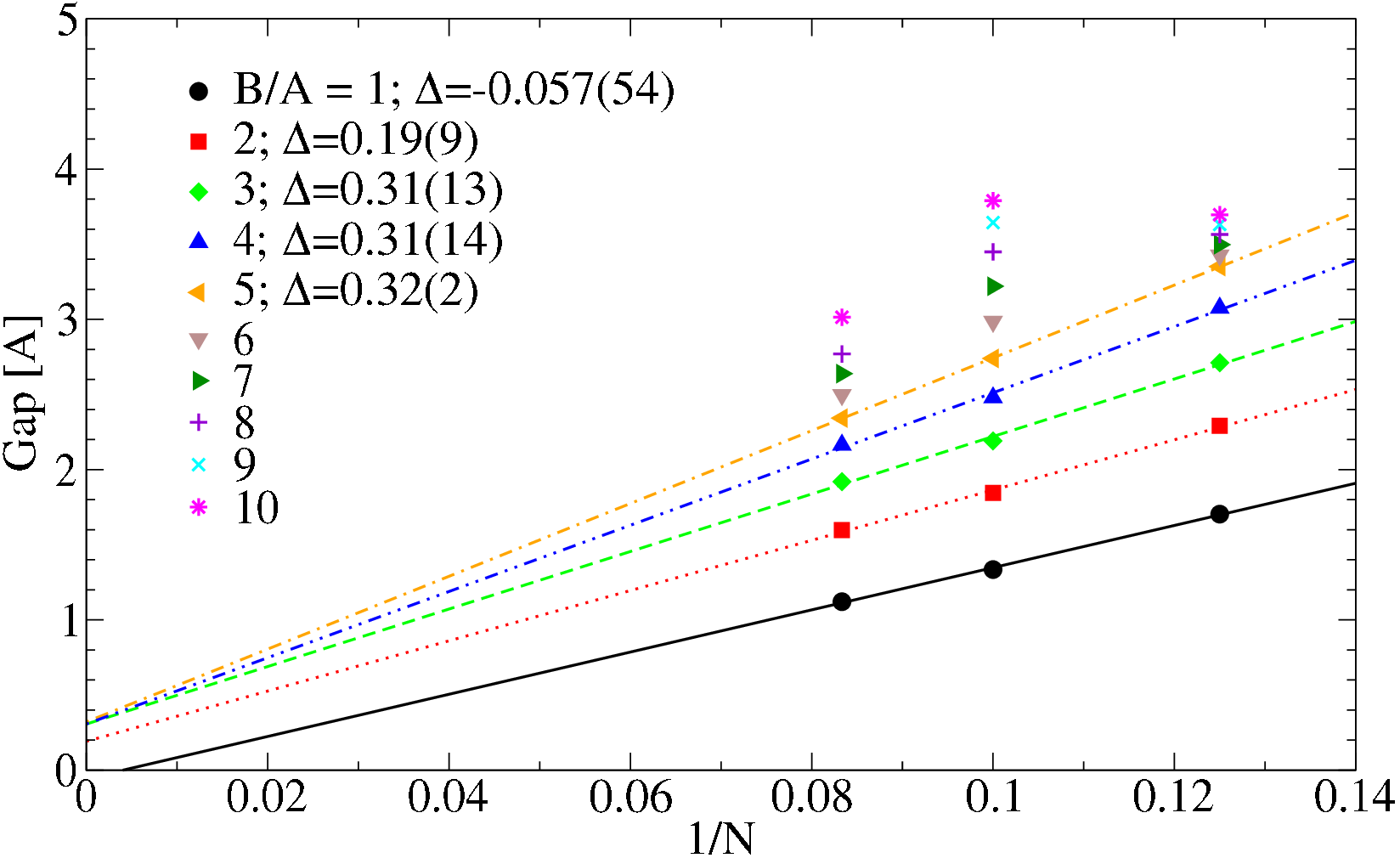}
\end{center}
\caption{\label{gapextra}
(Color online)
The gap of charged excitation in the Gaffnian model for various system sizes and parameter $B/A$ of the model interaction.
The value of the gaps in thermodynamic limit is given in those cases where a linear extrapolation is feasible.
}
\end{figure}

We believe that numerical studies can shed further light on this issue. There are two relevant pseudopotentials ($A$ and $B$) in the Gaffnian model, and the gap can be varied (at least for finite systems) by adjusting their relative strength, which gives a greater space of parameters in which to look for a state that has a gap in the thermodynamic limit.  Ref.~\onlinecite{unitary} considered the gap to creating a neutral excitation of the Gaffnian model (which is the lowest energy required to create an excitation at $2Q=5N/2-4$) and showed that it may vanish in the thermodynamic limit for $A=B$.
We give in Fig.~\ref{gapextra} the gap to creating a charged excitation, defined as
\begin{equation}
\frac{E\left(\frac{5N}{2}-3\right)+E\left(\frac{5N}{2}-5\right)-2E\left(\frac{5N}{2}-4\right)}{2}=\frac{E\left(\frac{5N}{2}-5\right)}{2}.
\end{equation}
where $E(2Q)$ is the ground state energy for the Gaffnain model at flux $2Q$.
The gap shows significant $N$ dependence.
A linear extrapolation of the gap appears satisfactory for $B/A\le 5$, and produces a nonzero gap for $3\lesssim B/A\lesssim 5$.
Notice, however, that this extrapolation is based on three data points only;
thus the system sizes accessible in our study do not allow a conclusive answer to the question of whether the gap survives in the thermodynamic limit.

If the gap vanishes in the thermodynamic limit for the Gaffnian model, then obviously the concept of braiding statistics is not meaningful in the thermodynamic limit.   However, if the gap vanishes as $1/N$ (as numerical calculations indicate),
it may be possible to define adiabatic braiding of quasiholes for a system that is sufficiently large but not infinite. In any case, for the remainder of this article, we will not address the issue of braiding statistics or the thermodynamic limit, but will only be concerned the Hilbert space counting in finite systems where the system is clearly gapped.

\section{Quasiholes}
\label{testing}

\begin{figure}[!htbp]
\begin{center}
\includegraphics[width=\columnwidth,keepaspectratio]{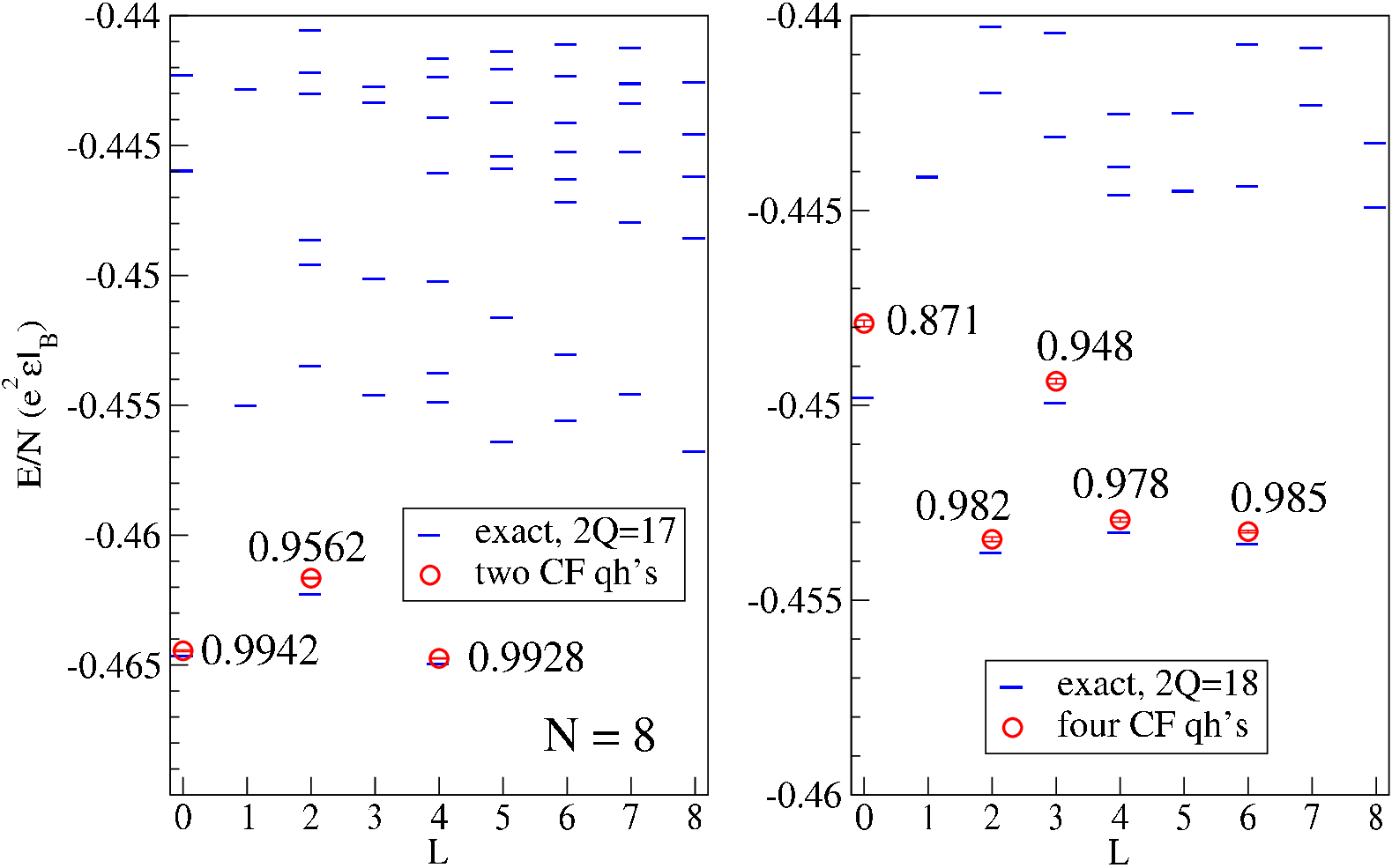}
\includegraphics[width=\columnwidth,keepaspectratio]{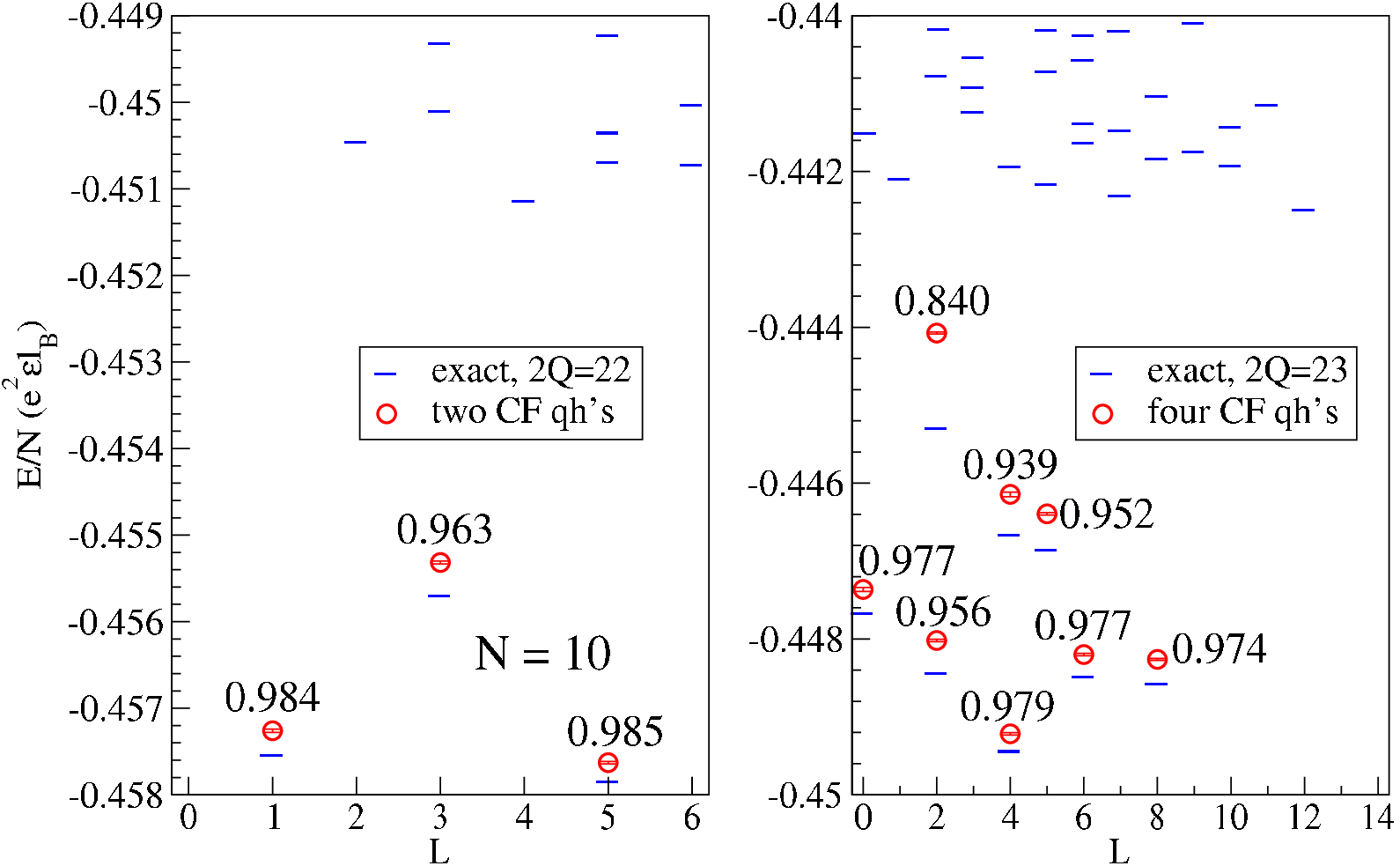}
\end{center}
\caption{\label{comparecf}
(Color online)
Comparison of the Coulomb energy of the composite fermion quasihole states with the exact low-energy spectrum with $N=8,10$ particles.
The numbers in the vicinity of CF states are squared overlaps with the corresponding exact state.
For ten particles we have used a Lanczos procedure for obtaining the low-energy states; only the low-energy part of the full Coulomb spectrum is shown.
}
\end{figure}

\begin{figure}[!htbp]
\begin{center}
\includegraphics[width=\columnwidth,keepaspectratio]{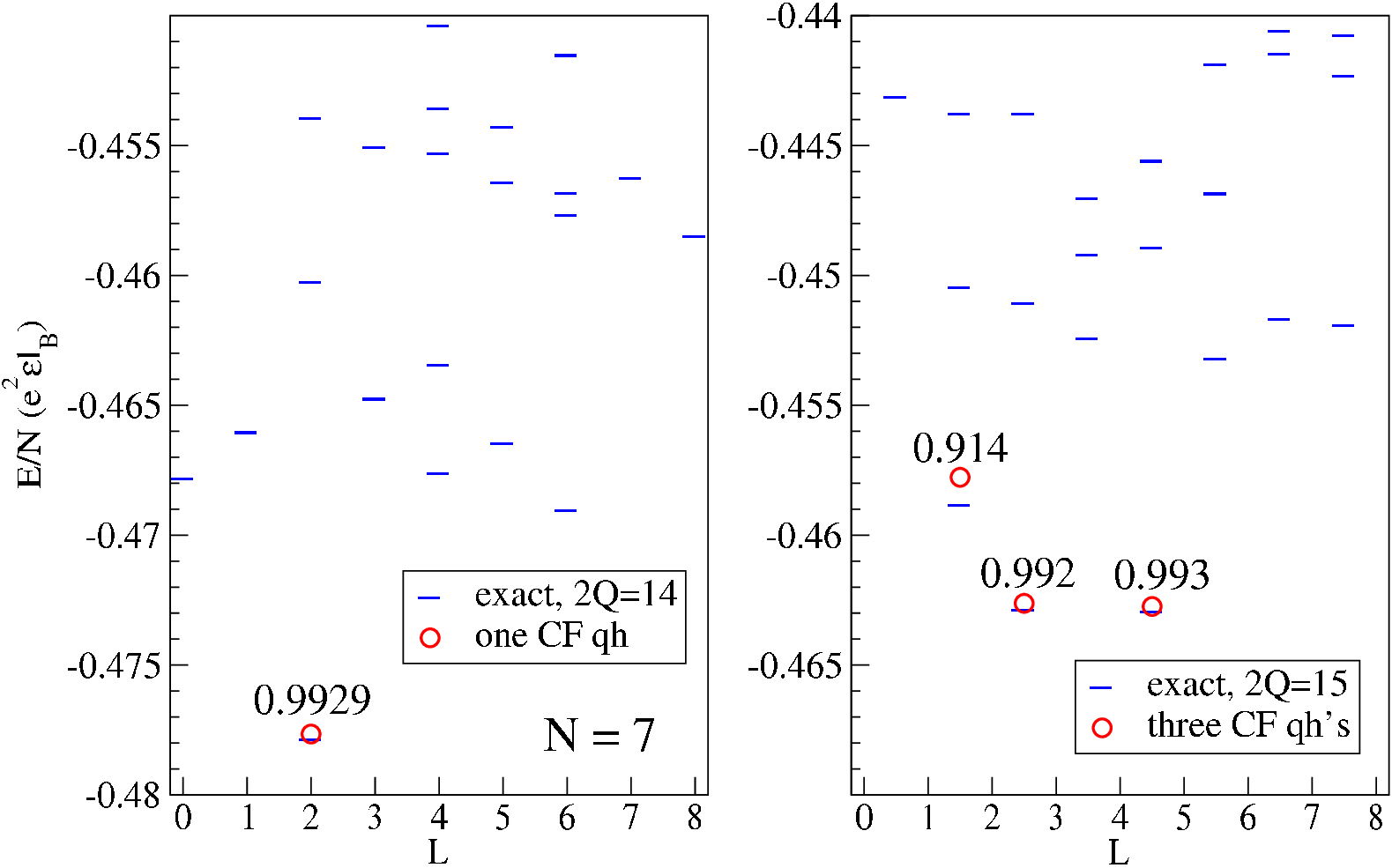}
\includegraphics[width=\columnwidth,keepaspectratio]{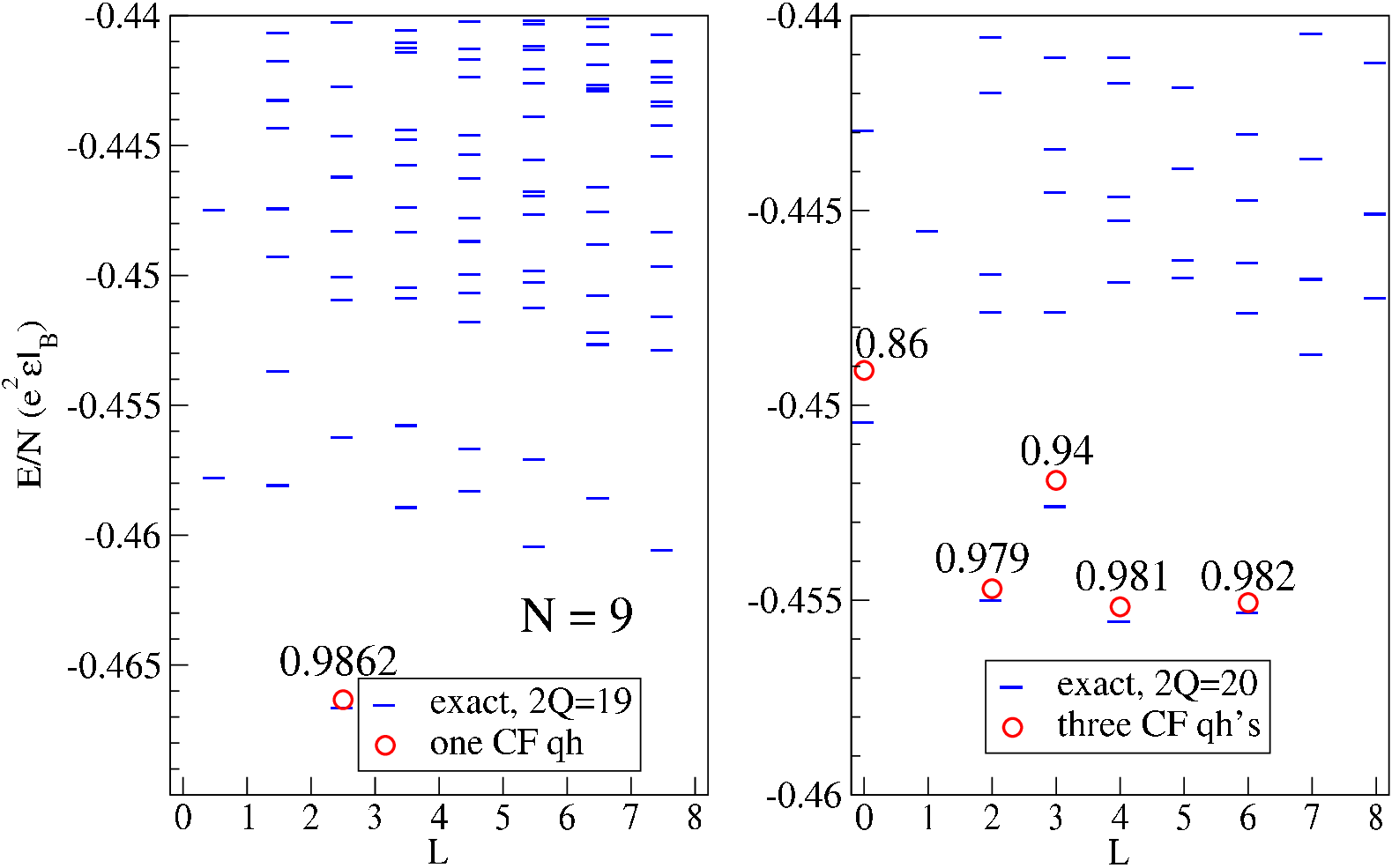}
\end{center}
\caption{\label{comparecfodd}
(Color online)
Comparison of the Coulomb energy of the composite fermion quasihole states with the exact low-energy spectrum with $N=7,9$ particles.
The numbers in the vicinity of CF states are squared overlaps with the corresponding exact state.
}
\end{figure}

As stated previously, the Gaffnian model gives a reasonably accurate approximation for the 2/5 ground state.  In this section we test it for quasiholes.  For contrast, a comparison between the CF model and the Coulomb solution is also given.

\subsection{CF-quasiholes}

Figure~\ref{comparecf} shows the exact Coulomb spectra for 8 and 10 particles (upper and lower panels, respectively) for two and four quasiholes (left and right panels, respectively).  Figure~\ref{comparecfodd} shows the exact Coulomb spectra for 7 and 9 particles (upper and lower panels, respectively) for one and three quasiholes (left and right panels, respectively). 

Of interest here is, that some states break off from the others to form a low-energy band, which we call the Coulomb quasihole band.  
The quantum numbers of states in the low-energy band in these figures are in complete agreement with those predicted by the CF theory (Tables \ref{counting} and \ref{countingodd}).  Figs.~\ref{comparecf} and \ref{comparecfodd}  also show a comparison between the exact Coulomb energies and the CF energies, as well as the overlaps between the exact Coulomb states and the CF wave functions.   (For four quasiholes at 10 particles, CF diagonalization is needed to produce the energies and the wave functions, because there are two states each at total orbital angular momentum values of $L=2$ and 4, but for all other cases the CF theory provides a unique state.) Evidently, the CF theory gives an excellent qualitative and quantitative account of the quasiholes.

\subsection{Gaffnian quasiholes}

The Gaffnian and CF models predict the same quantum numbers for the one, two, or three quasihole bands, which are therefore also in agreement with the low-energy Coulomb band.
For four quasiholes, in contrast, the Gaffnian model predicts significantly more states (Table \ref{counting}) than observed (Fig.~\ref{comparecf}), indicating a lack of one-to-one correspondence between the Gaffnian and the Coulomb quasiholes.

\begin{figure}[!htbp]
\begin{center}
\includegraphics[width=\columnwidth,keepaspectratio]{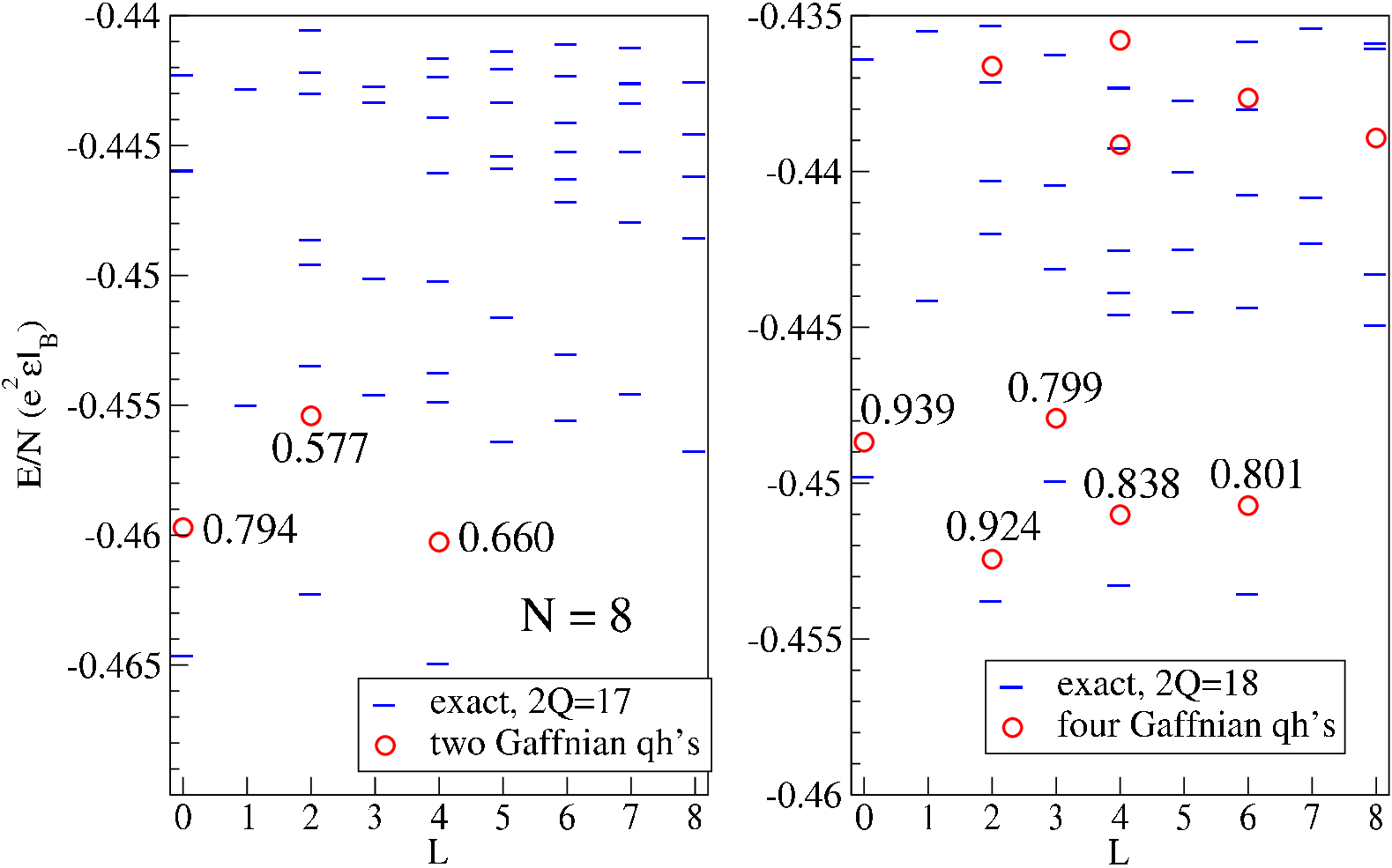}
\includegraphics[width=\columnwidth,keepaspectratio]{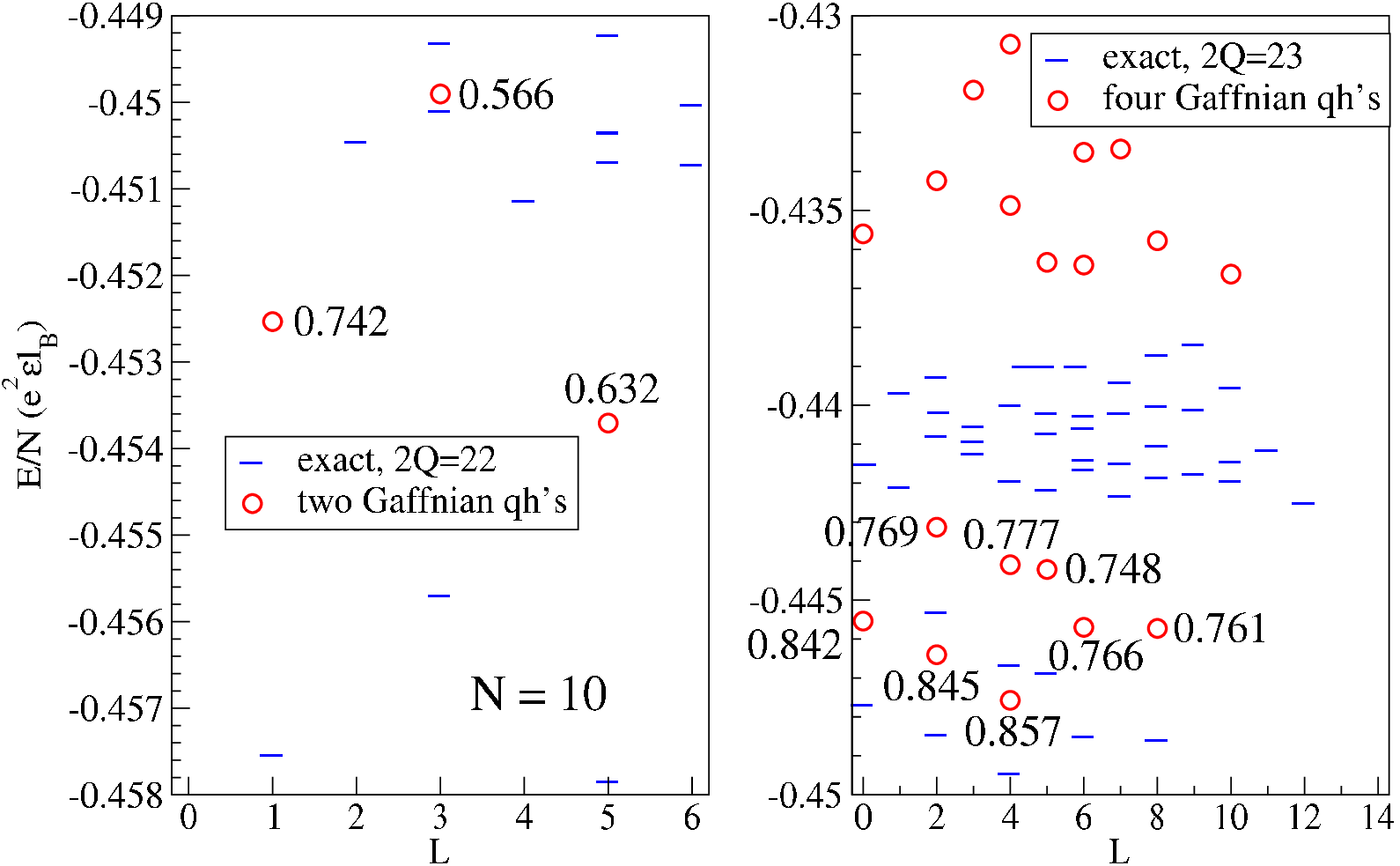}
\end{center}
\caption{\label{comparegaffnian}
(Color online)
Comparison of the Coulomb energy of the Gaffnian quasihole states with the exact low-energy spectrum with $N=8,10$ particles.
The numbers in the vicinity of Gaffnian states are squared overlaps with the corresponding exact state.  For four quasiholes the overlaps are given only for those states that qualitatively follow the lowest band of Coulomb states.
For ten particles we have used a Lanczos procedure for obtaining the low-energy states; only the low-energy part of the full Coulomb spectrum is shown.
}
\end{figure}

\begin{figure}[!htbp]
\begin{center}
\includegraphics[width=\columnwidth,keepaspectratio]{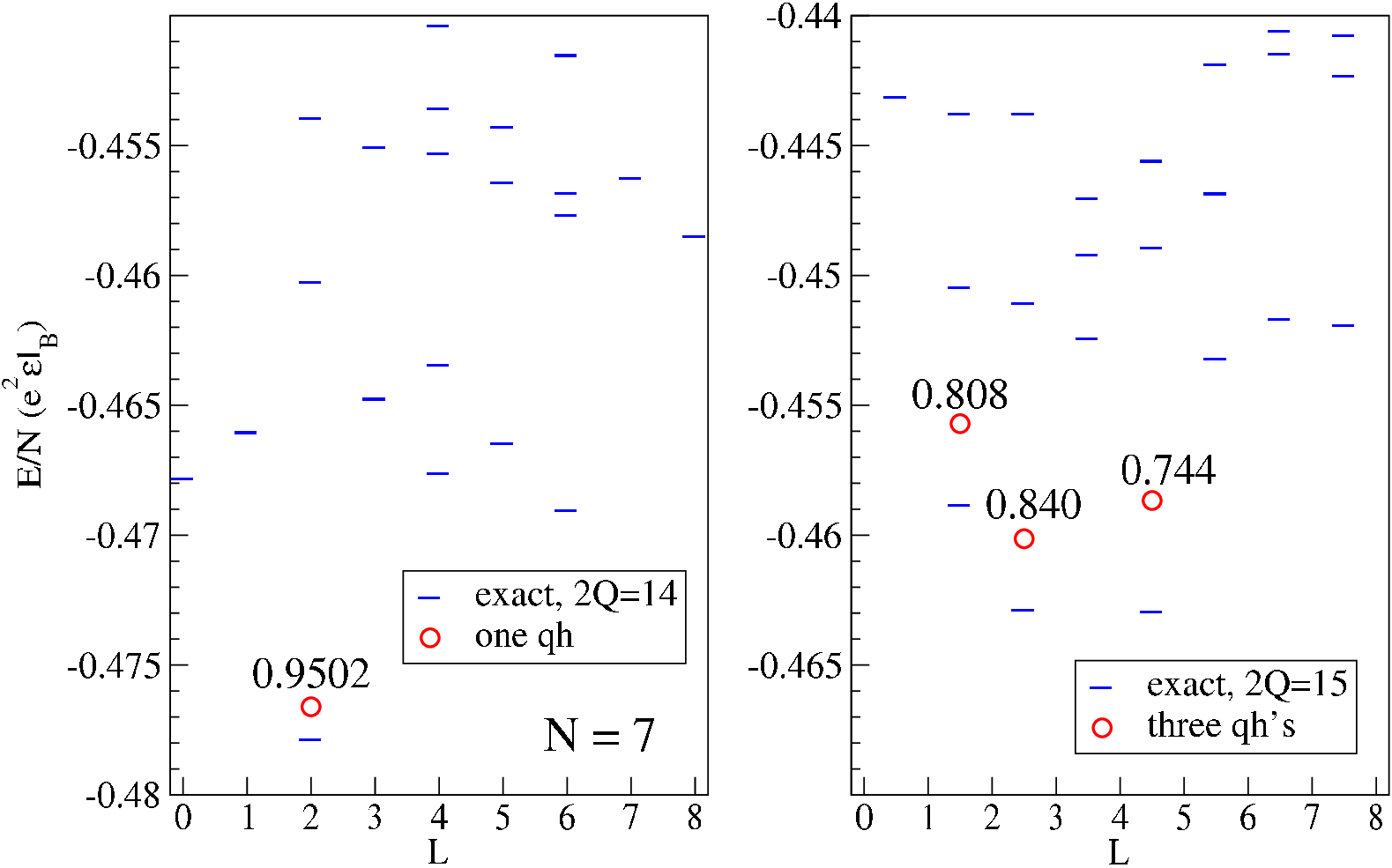}
\includegraphics[width=\columnwidth,keepaspectratio]{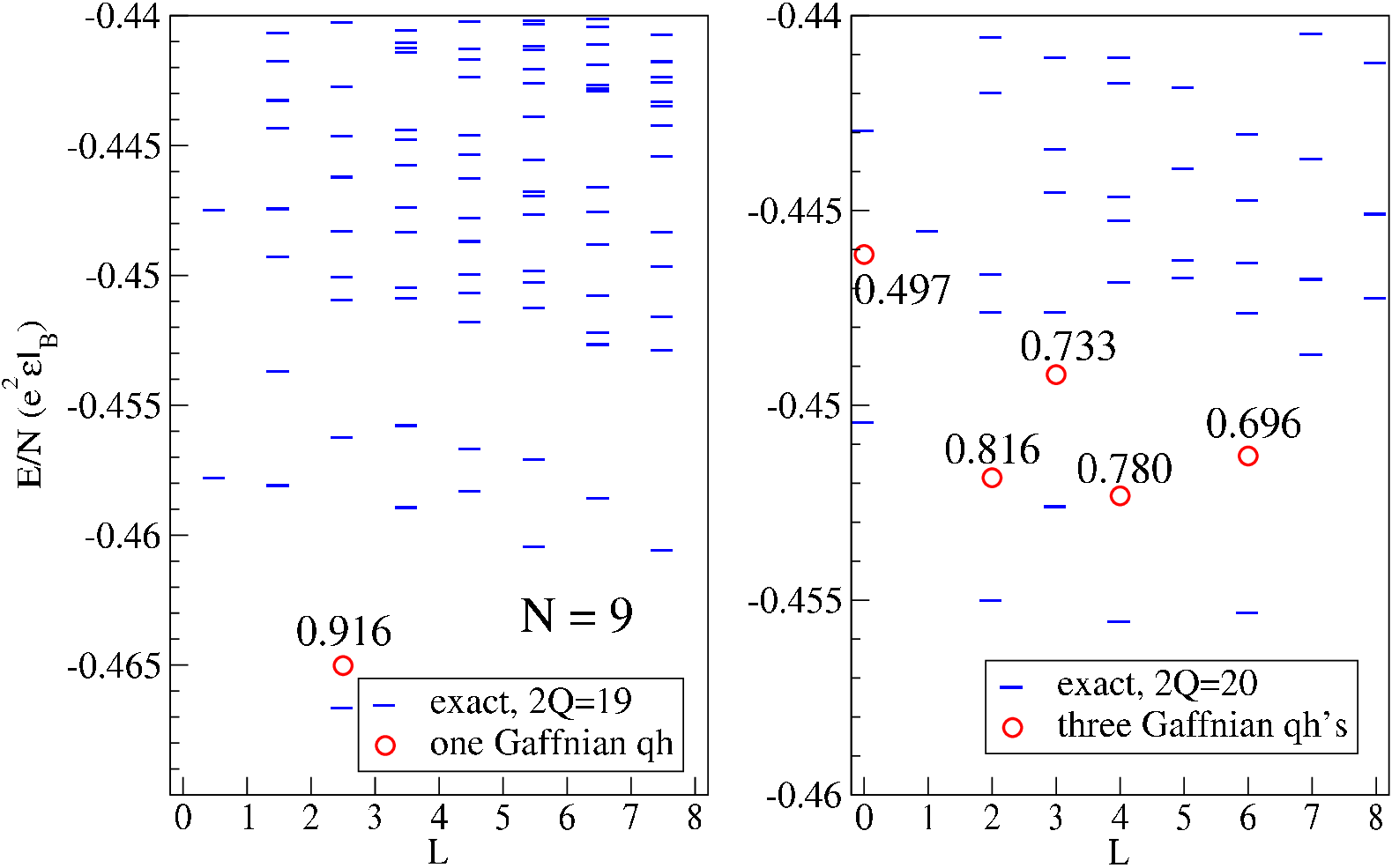}
\end{center}
\caption{\label{comparegaffnianodd}
(Color online)
Comparison of the Coulomb energy of the Gaffnian quasihole states with the exact low-energy spectrum with $N=7,9$ particles.
The numbers in the vicinity of Gaffnian states are squared overlaps with the corresponding exact state.
}
\end{figure}

The overlaps between the Gaffnian quasihole wave functions and the exact Coulomb wave functions for one to three quasihole states are shown in Figs.~\ref{comparegaffnian} and \ref{comparegaffnianodd}, and for four quasiholes in Table~\ref{fourqhoverlaps}, which gives the {\em cumulative} overlap between the {\em full} Gaffnian basis and an equal number of lowest Coulomb states, as defined in the caption. 
The overlaps for the Gaffnian quasiholes decay more rapidly compared to the CF-quasiholes as one moves away from the incompressible state.

Figures~\ref{comparegaffnian} and \ref{comparegaffnianodd} also show a comparison between the Gaffnian and the exact energies.  The Gaffnian spectrum is obtained by  diagonalizing the Coulomb Hamiltonian within the
basis of the Gaffnian quasihole wave functions.  
(The overlaps shown in Fig.~\ref{comparegaffnian} for four quasiholes only compare a small subset of the Gaffnian basis, as picked out by the Coulomb interaction, with the Coulomb quasiholes, and are thus not a test of the original Gaffnian model; they are further discussed in Sec.~\ref{contain}.)

Strictly within the Gaffnian model itself, there would be no logical way to rule out that the lack of one-to-one correspondence between the Gaffnain and the Coulomb solutions is a finite size artifact, and that an agreement would be obtained for sufficiently large systems. However, a comparison with the CF theory suggests that the mismatch between the {\em full} Gaffnian and the Coulomb quasihole sectors is {\em not} a finite size effect.  Given that the Gaffnian and CF-quasihole spectra are different, the plausible assumption that the Coulomb solution for four quasiholes matches with the CF theory excludes a similar matching with the Gaffnian model no matter how large the system size.

\begin{table}[htb]
\begin{center}
\begin{tabular}{c|c|c|c|c|c|c|c|c|c}
\hline\hline
$N$ &  $L=0$ & 2     & 3     & 4     & 5     & 6     & 7     & 8     & 10 \\
\hline
6  & 0.940   & 0.930 & 0.262 & 0.685 & -     & 0.620 & -     & -     & -\\
8  & 0.470   & 0.356 & 0.799 & 0.613 & 0.205 & 0.674 & -     & 0.465 & -\\
10 & 0.670   & 0.490 & 0.004 & 0.499 & 0.682 & 0.490 & 0.003 & 0.560 & 0.467 \\
\hline\hline
\end{tabular}
\end{center}
\caption{\label{fourqhoverlaps}Cumulative squared overlaps between the Gaffnian four quasihole sector and the lowest energy states for Coulomb interaction at $2Q=5N/2-2$.
The overlap at a given $L$ is defined as ${\cal O}=\sum_{i,j}^{\cal N}|\langle\Psi^{\text{G}}_{\rm 4-qh, i}|\Psi^{\rm C}_{\rm 4-qh, j}\rangle|^2 / {\cal N}$, where
${\cal N}$ is the number of degenerate multiplets of $\hat H^{\text{G}}$ at $L$ (Table \ref{counting}), and $i,j=1,\cdots, {\cal N}$.
The states $\Psi^{\rm C}_{\rm 4-qh, j}$ represent the ${\cal N}$ lowest energy eigenstates of the Coulomb interaction.
The overlaps for two quasiholes are shown in left panels of Fig.~\ref{comparegaffnian}, for one and three quasiholes in Fig.~\ref{comparegaffnianodd}.
}
\end{table}

\begin{figure}[!htbp]
\begin{center}
\includegraphics[width=\columnwidth,keepaspectratio]{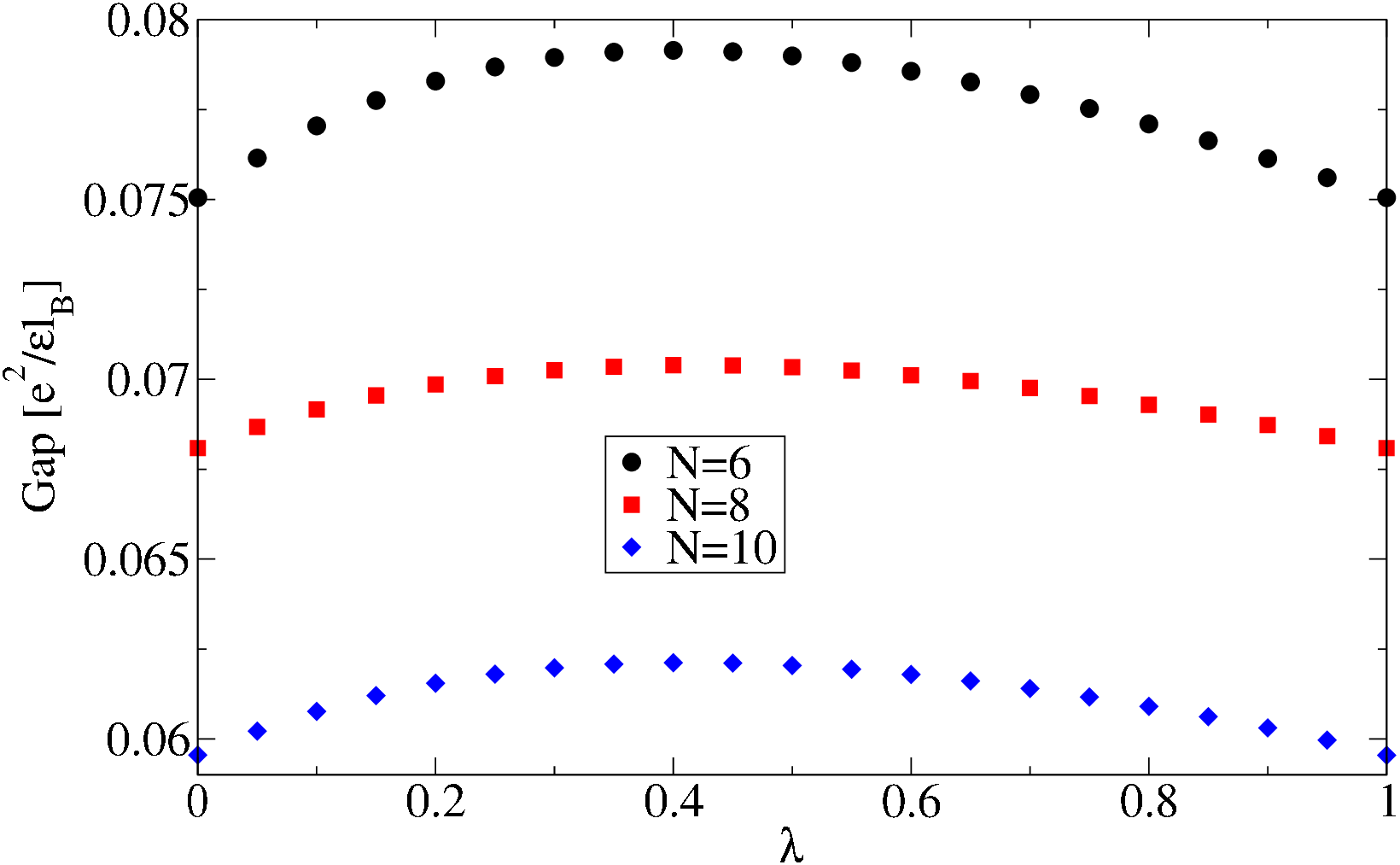}
\end{center}
\caption{\label{gapfig}
(Color online)
The gap for the mixed interaction $H_\lambda$ for $N=6,8,10$ particles as a function of $\lambda$.  In this figure as well as in Figs. \ref{gfoverlaps} and \ref{gfoverlapsodd} we have taken $A=B$ in the Gaffnain interaction, and fixed its value by the requirement that the gap for neutral excitation be equal for $\lambda=1$ and $\lambda=0$.
}
\end{figure}

To further explore the relation between the Gaffnian and the Coulomb models, we study
the crossover from the Gaffnian state to the CF state by considering the interaction
\begin{equation}
\label{interpol}
\hat H_\lambda=\lambda \hat H^{\text{G}} + (1-\lambda)\hat H^{\text{C}}
\end{equation}
that interpolates between the Gaffnian model interaction $\hat H^{\text{G}}$ and the Coulomb interaction,
\[
\hat H^{\text{C}}=\sum_{i<j}\frac{1}{|r_i-r_j|}.
\]
There is an ambiguity regarding how to fix the relative energy scales of the Gaffnian and the Coulomb terms, because they do not have the same parametric dependences. We set $A=B$ for simplicity, and fix this constant by the requirement that the gap for neutral excitation be equal\cite{Storni} for $\lambda=1$ and $\lambda=0$ (for the Coulomb interaction, the gap corresponds to the CF roton energy).\cite{roton} As seen in Fig.~\ref{gapfig}, the gap never closes when $\lambda$ is tuned between the two extremes; in fact, it becomes stronger in the intermediate region.  This study shows that, at least for finite systems studied here, the Gaffnian state is smoothly connected to the Coulomb / CF ground state, and strongly suggests that this should be the case also in the thermodynamic limit {\em provided} the Gaffnian model produces an incompressible state in that limit.

\begin{figure}[!htbp]
\begin{center}
\includegraphics[width=\columnwidth,keepaspectratio]{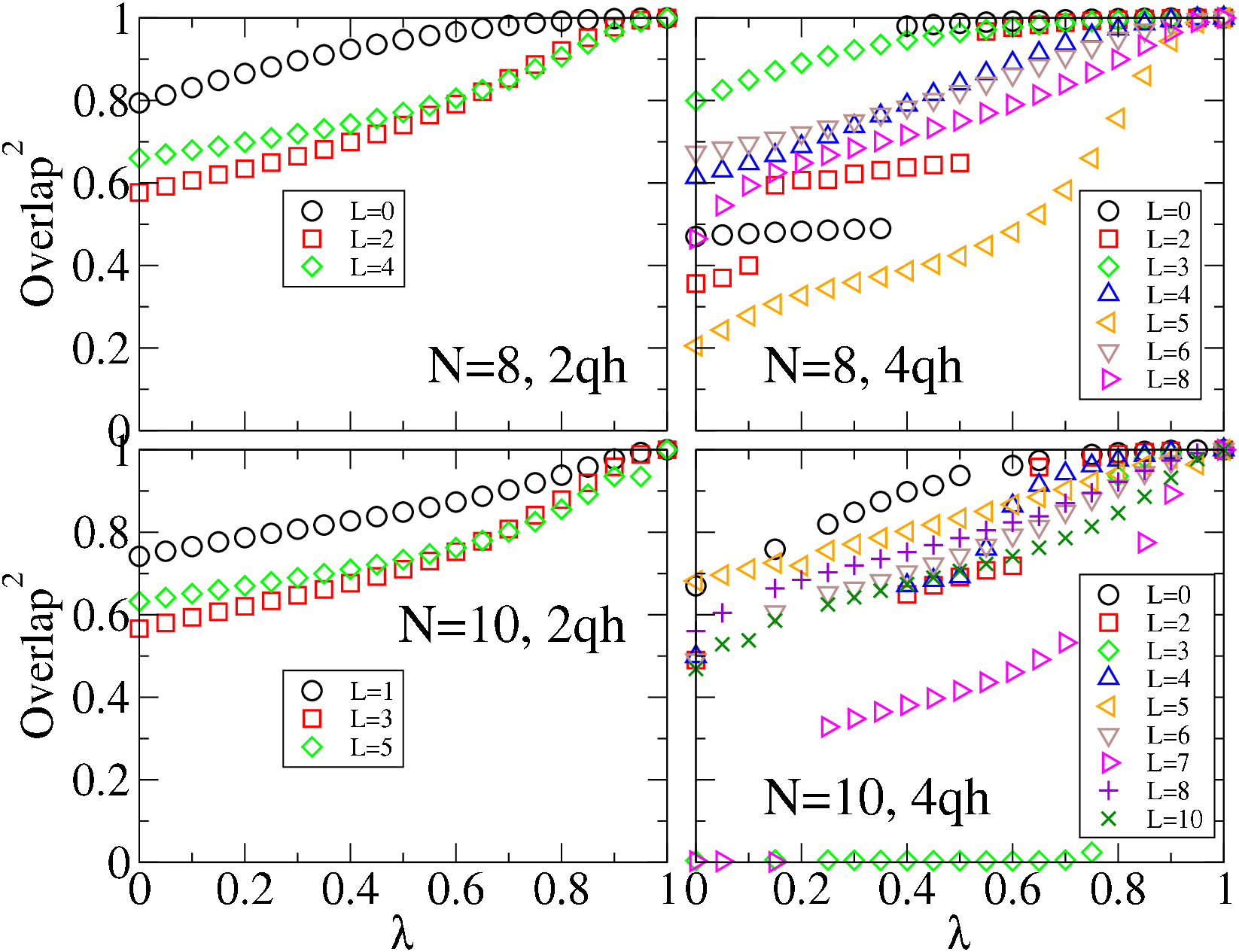}
\end{center}
\caption{\label{gfoverlaps}
(Color online)
The squared overlaps with the zero-energy branch of the Gaffnian interaction $H^{\text{G}}$ with the exact eigenstates of the model in Eq.~(\ref{interpol}), which interpolates between Coulomb and Gaffnian, for two and four quasiholes.  The results are shown for $N=8,10$ particles as a function of parameter $\lambda$; the interaction is pure Coulomb at $\lambda=0$ and pure Gaffnian at $\lambda=1$. The overlaps for the four quasihole sector are
defined in the same manner as in Table \ref{fourqhoverlaps}.
A few points have been omitted where the L\'anczos method did not clearly resolve  nearly degenerate states.
}
\end{figure}

\begin{figure}[!htbp]
\begin{center}
\includegraphics[width=\columnwidth,keepaspectratio]{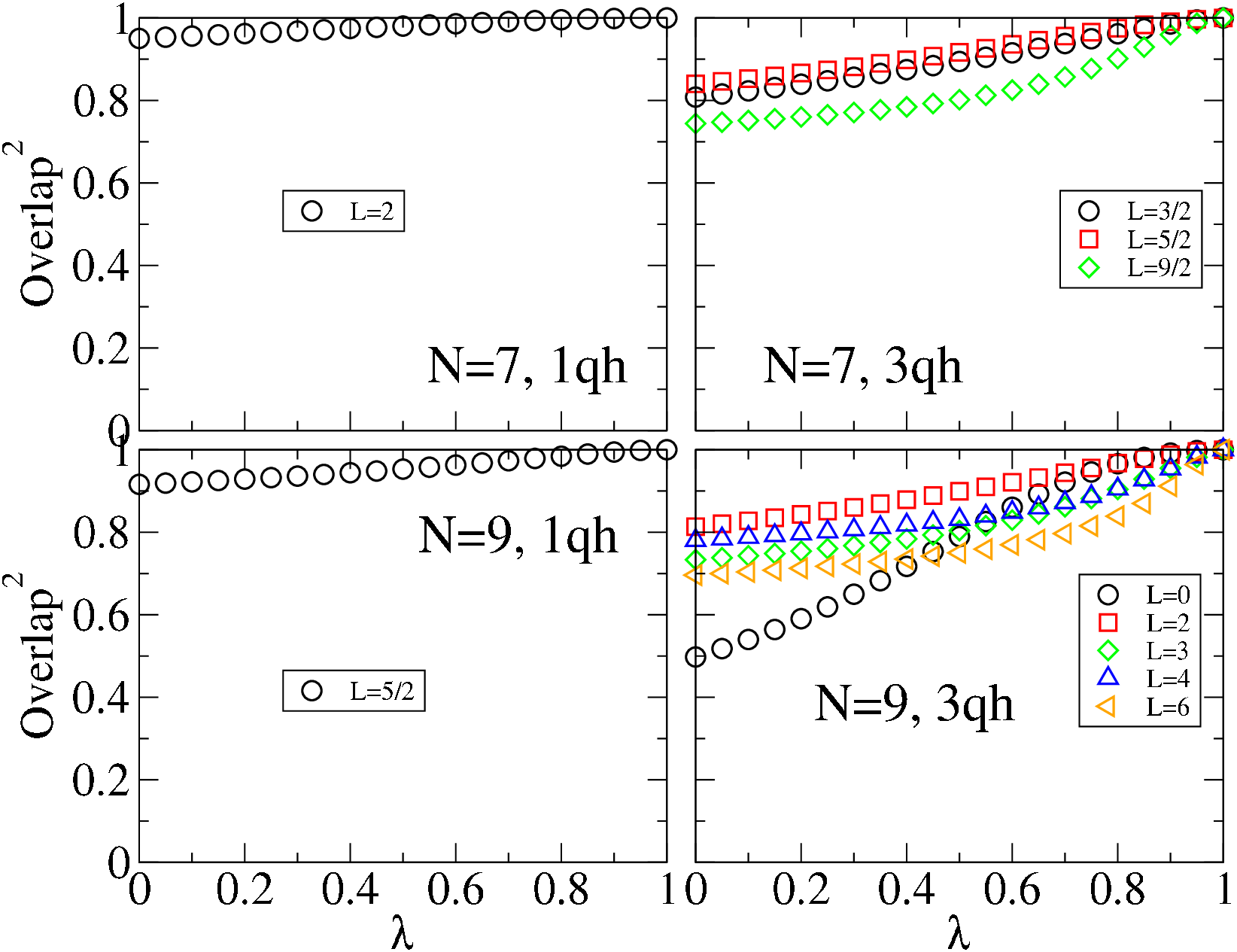}
\end{center}
\caption{\label{gfoverlapsodd}
(Color online)
The squared overlaps with the quasihole branch of the Gaffnian interaction $H^{\text{G}}$ with the exact eigenstates of the model in Eq.~(\ref{interpol}), which interpolates between Coulomb and Gaffnian, for one and three quasiholes.  The results are shown for $N=7,9$ particles as a function of parameter $\lambda$; the interaction is pure Coulomb at $\lambda=0$ and pure Gaffnian at $\lambda=1$. Notice there is a unique state for each angular momentum, and that the counting by the Gaffnian
and the CF models coincide.
}
\end{figure}

Figures~\ref{gfoverlaps} and \ref{gfoverlapsodd} show the evolution of the Gaffnian quasihole states as the interaction is tuned from Gaffnian ($\lambda=1$ point) to Coulomb ($\lambda=0$ point).  The cumulative overlaps of the Gaffnian quasihole band (that is, all states with zero-energy at the Gaffnian point) are evaluated with the corresponding lowest energy eigenstates of the model in Eq.~(\ref{interpol}).
For one, two, or three quasiholes, the overlaps decay as one moves from $\lambda =1$ to $\lambda=0$, and the behavior is continuous.  For four quasiholes, the overlaps decay more rapidly, but with discontinuous jumps as a function of $\lambda$. These discontinuous jumps are an indication of level crossings in the actual spectra as a function of $\lambda$. As discussed below in the context of Fig.~\ref{comparegaffnian}, some of the Gaffnian quasihole states are pushed to very high energies for the Coulomb problem forming an ``upper subband."

It is not possible to locate the precise parameter range where the Gaffnian model is valid from our small systems studies. However, the range near $\lambda=1$
where all of the Gaffnian quasihole states have a high overlap (by standards of Table \ref{overlap}) with the exact quasihole states appears quite narrow.

\section{Does the Gaffnian model contain composite fermions?}
\label{contain}

The observation that the Gaffnian quasihole space is larger than the CF-quasihole space raises the question if the latter is contained in the former.  That is certainly not ruled out a priori.
For reasons discussed below, we believe that the answer to the question is in the affirmative, at least in a qualitative sense.

A noticeable aspect of Fig.~\ref{comparegaffnian} is, that the Coulomb interaction splits the four Gaffnian quasihole sector into two sub-bands.
The upper sub-band of the Gaffnian quasihole states is pushed up into the ``continuum'' of high energy Coulomb states, but the lower energy sub-band has a one-to-one correspondence with the CF-quasihole band. That gives an indication that at least the qualitative physics of composite fermions is contained in the Gaffnian quasihole basis (although it is brought out only by a diagonalization of the Coulomb interaction).

To ascertain the quantitative extent to which the CF-quasihole wave functions can be accommodated within the Gaffnian quasihole sector, we give in Tables \ref{twoqhcontained} and \ref{fourqhcontained} the relevant overlaps for two and four quasiholes.  We first note that for two quasiholes, where the Gaffnian and the CF-quasihole sectors contain the same number of states, the overlaps between them are significantly smaller than those between the 2/5 Gaffnian and CF ground states (Table \ref{overlap}).  For four quasiholes, one might have expected the overlaps in Table \ref{fourqhcontained} to {\em increase} with $N$, given that the size of the Gaffnain quasihole sector grows much more rapidly with the number of quasiholes than the size of the CF-quasihole sector, thus allowing for greater flexibility.  However, the overlaps do not increase (at least substantially) for the systems studied either for two or four quasiholes.  This seems surprising at first, but can be understood from the fact that the upper band lies at very high energies, indicating that many of the Gaffnain quasihole basis states are practically  orthogonal to the CF-quasihole states.
The picture that seems to emerge from these observations is that while the overlaps between the Gaffnian and the CF-quasihole spaces do not increase as we increase the system size for a fixed number of quasiholes, they do increase as we increase the number of quasiholes.  As shown in Ref.~\onlinecite{Gaffnian}, the degeneracy of the Gaffnian quasiholes has two parts, one associated with the positional degeneracy and the other with the degeneracy of the zero modes. The latter gives the degeneracy when the positions of the quasiholes are fixed, and depends only on the number of quasiholes, not on the system size $N$. This suggests that the extent to which the Gaffnian quasihole band contains the {\em quantitative} correlations built in the CF state is related to the degeneracy of the zero modes.

In any case, the important point is that while the zero-energy Gaffnian quasihole band does not have a one-to-one correspondence with the low-energy band in the Coulomb spectrum,  the Gaffnian quasihole band is split by the Coulomb interaction to produce a lower energy band that does match the Coulomb band. The Gaffnian model thus becomes unstable to the formation of composite fermions when the two-body Coulomb interaction is switched on, thus altering the character of the ``physical quasiparticles."

\begin{table}[htb]
\begin{center}
\begin{tabular}{c|c|c|c|c|c|c}
\hline\hline
$N$ &  $L=0$ & 1 & 2 & 3 & 4 & 5  \\
\hline
6 & - &  0.777(2) & - & 0.696(2) & - & - \\
8 & 0.818(1) & - & 0.701(2) & - & 0.653(1) & - \\
10 & - & 0.769(5) & - & 0.676(6) & - & 0.621(6) \\
\hline\hline
\end{tabular}
\end{center}
\caption{\label{twoqhcontained}
Squared overlaps $|\langle\Psi^{\text{G}}_{\rm 2-qh, j}|\Psi^{\rm CF}_{\rm 2-qh, i}\rangle|^2$ showing the extent to
which the CF states are contained in the Gaffnian quasihole sector at $2Q=5N/2-3$ (two quasiholes).
}
\end{table}

\begin{table}[htb]
\begin{center}
\begin{tabular}{c|c|c|c|c|c|c|c}
\hline\hline
$N$ &  $L=0$   & 2        & 3     & 4        & 5     & 6     & 8 \\
\hline
6  & 0.939(10) & 0.926(5) & -     & 0.906(2) & -     & -        & - \\
8  & 0.984(6)  & 0.896(3) & 0.921 & 0.872(2) & -     & 0.880(2) & - \\
10 & 0.857(8)  & 0.85(1)  & -     & 0.843(4) & 0.845(4) & 0.807(8) & 0.823(2) \\
   &           & 0.876(8) &       & 0.862(6) &        &        &   \\
\hline\hline
\end{tabular}
\end{center}
\caption{\label{fourqhcontained}
Squared overlaps showing the extent to which the CF states are contained in the Gaffnian quasihole sector at $2Q=5N/2-2$ (four quasiholes).
The overlap for a CF state $i$ with angular momentum $L$ is defined as
${\cal O}_i=\sum_{j}^{\cal N}|\langle\Psi^{\text{G}}_{\rm 4-qh, j}|\Psi^{\rm CF}_{\rm 4-qh, i}\rangle|^2$, where
${\cal N}$ is the number of degenerate multiplets of $\hat H^{\text{G}}$ at $L$ (Table \ref{counting}).
}
\end{table}

\section{Discussion}
\label{discussion}

The character change of the excitations in this manner might at first seem surprising, but, with further thought, it is actually to be anticipated.  Essentially, whenever we have a vastly degenerate set of states, turning on even a slight perturbation can have nonperturbative consequences. It is instructive to review some other known examples in the context of the FQHE, which strongly suggest that a Hilbert space reduction / rearrangement is not just possible, but is very likely to be generic.

(i) The first example is the FQHE itself. For noninteracting electrons, there is a vast degeneracy of many body ground states in the lowest Landau level.  The introduction of an arbitrarily weak repulsive interaction creates composite fermions and the lowest Landau level splits into their $\Lambda$ levels, which reduces the degeneracy by opening new gaps in the spectrum.

(ii) The hard-core model interaction $V_1=\sum_{i<j}\nabla^2_i\delta^{(2)}(z_i-z_j)$, which acts only upon two electrons with relative angular momentum one, produces a large number of degenerate states of the form
\begin{equation}
\prod_{j<k}(z_j-z_k)^3 F_{\rm S}[\{ z_i\}]
\end{equation}
for $\nu<1/3$, where $F_{\rm S}[\{ z_i\}]$ is a symmetric polynomial of the electron coordinates.  The degeneracy is equal to the number of partitions of $M$, where $M$ is the number of additional flux quanta relative to the state at $1/3$. Clearly, this does not represent the correct physics, as is most obvious from the fact that it misses a large number of FQHE states with $\nu<1/3$, such as that at 2/7.

A different model for this region is in terms of composite fermions carrying four inverse vortices (also known as reverse flux attachment).\cite{Wu93} The wave function is given by
\begin{equation}
\mathcal P_{\text{LLL}}\prod_{j<k}(z_j-z_k)^4 \Phi_{\nu^*}^*,
\end{equation}
where $\nu^*=1$ gives a FQHE state at 1/3 and $\nu^*=2$ gives a FQHE state at 2/7. [$\nu^*=n$ corresponds to $n/(4n-1)$.]
The counting of states from this prescription as the filling factor is changed from 1/3 to 1/4 is in one-to-one correspondence with the counting from filling factor $\nu^*=1$ to $\nu^*=\infty$, which is in general much smaller than that predicted by the $V_1$ model, and, in particular, gives unique ground state at $n/(4n-1)$.  From the perspective of the $V_1$ model, the residual interaction causes a nonperturbative rearrangement of the low-energy states.  Thus, even though the 1/3 state evolves continuously as we go from $V_1$ to the full Coulomb interaction, the structure on the quasihole side changes in a qualitative manner. (Even lower filling factors can be similarly understood in terms of composite fermions with appropriate number of vortices attached to them.) It is worth noting that the CF states (which are essentially exact) are contained very accurately within the subspace of states that have zero-energy for the $V_1$ interaction.

(iii) Consider composite fermion states
\begin{equation}
\prod_{j<k}(z_j-z_k)^2 \Phi_{\nu^*},
\end{equation}
for a nonintegral $\nu^*$.  The model of noninteracting composite fermions predicts a band of quasi-degenerate states, whose dimensions can be determined straightforwardly by analogy to IQHE at $\nu^*$.  However, it is likely that any residual interaction between composite fermions
in the partially filled $\Lambda$ level will cause a rearrangement of the states, thereby further reducing the dimension of the low-energy Hilbert space.  For example, for certain values of $\nu^*$ the CF-quasiparticles can arrange themselves into a crystal or stripes;\cite{stripes} and for certain other values of $\nu^*$ they can form their own FQHE state\cite{theory4p11} (which produces fractions\cite{exp4p11} such as 4/11) for which the quasiparticles are very different from those of 1/3 or 2/5. This is again an example where the low-energy Hilbert space is qualitatively altered due to the weak residual interaction between the quasiparticles.  Note that the rearrangement can possibly occur entirely within the partially filled $\Lambda$ level, without closing the $\Lambda$ level gap.

It should be noted that the phase transitions discussed in these examples, due to the turning on of the ``rest of the interaction," are ``topological," beccause the new state is described in terms of composite fermions carrying a different number of vortices; also the new quasiparticles have different fractional local charge and braiding statistics.

In summary, we have examined the Gaffnian model for the 2/5 FQHE and found that, while it gives a reasonably accurate wave function for the 2/5 Coulomb ground state (at least for small systems), it is inadequate for quasiholes.  This study has general implications for FQHE beyond the 2/5 state.  Most importantly, it shows, not surprisingly, that a consideration of the ground state alone is insufficient for the demonstration of the validity of a model; it is necessary to test it for excitations as well, because two ground state wave functions that a have high overlap can have qualitatively distinct quasiparticles.

We thank Vadim Cheianov, Nicolas Regnault, Steve Simon and Arek W\'{o}js for useful discussions.
C.\ T.\ received support from the Lancaster University-EPSRC Portfolio Partnership.
Computations were performed on the High Performance Cluster at Lancaster University.

\newcommand{\PRL}{Phys.\ Rev.\ Lett.}
\newcommand{\PRB}{Phys.\ Rev.\ B}
\newcommand{\PRD}{Phys.\ Rev.\ D}
\newcommand{\NPB}{Nucl.\ Phys.\ B}

\end{document}